\documentclass[11pt,a4paper]{article}
\usepackage{graphicx}

\title{Retardation Effects in Gravitation}
\author{
Asher Yahalom$^{a,b}$\\
$^a$ Ariel University, Kiryat Hamada POB 3, Ariel 40700, Israel\\
$^b$ Princeton University, Princeton, New Jersey 08543, USA\\
e-mail:  asya@ariel.ac.il}

\begin{document}
\maketitle

\renewcommand{\thepart}{\Large \arabic{part}}
\newcommand{\beq} {\begin{equation}}
\newcommand{\enq} {\end{equation}}
\newcommand{\ber} {\begin {eqnarray}}
\newcommand{\enr} {\end {eqnarray}}
\newcommand{\eq} {equation}
\newcommand{\eqn} {equation }
\newcommand{\eqs} {equations }
\newcommand{\ens} {equations}
\newcommand {\er}[1] {equation (\ref{#1}) }
\newcommand {\ern}[1] {equation (\ref{#1})}
\newcommand {\ers}[1] {equations (\ref{#1})}
\newcommand {\Er}[1] {Equation (\ref{#1}) }
\newcommand {\Ern}[1] {Equation (\ref{#1})}
\newcommand{\mn}  {{\mu \nu}}
\newcommand{\sn}  {{\sigma \nu}}
\newcommand{\rhm}  {{\rho \mu}}
\newcommand{\sr}  {{\sigma \rho}}
\newcommand{\ab}  {{\alpha \beta}}
\newcommand{\bh}  {{\bar h}}
\newcommand{\br}  {{\bar r}}
\newcommand{\citet} {\cite }
\newcommand{\citep} {\cite }
\renewcommand{\Box} {  }
\newcommand{\citeauthoryear} { }
\newcommand{\hdz}  {\frac{1}{2} \Delta z}


\pagestyle{myheadings}

\begin {abstract}

Galaxies are huge physical systems having dimensions of many tens of thousands of light years. Thus
any change at the galactic center will be noticed at the rim only tens of thousands of years later.
Those retardation effects seems to be neglected in present day galactic modelling used to calculate rotational velocities of matter in the rims of the galaxy and surrounding gas. The significant differences between the predictions of Newtonian instantaneous action at a distance and observed velocities are usually explained by either assuming dark matter or by modifying the laws of gravity (MOND). In this paper we will show that taking general relativity seriously without neglecting retardation effects one can explain the radial velocities of galactic matter without postulating dark matter.

\end {abstract}

\section {Introduction}

\begin{figure}
\includegraphics[width=\columnwidth]{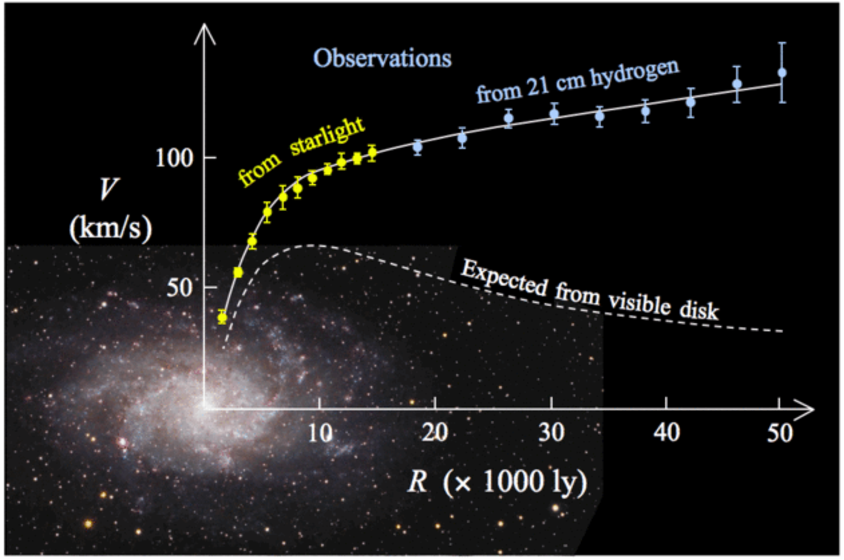}
\caption{M33 rotation curve \citet{Corbelli}}
\label{M33rc}
\end{figure}

The general theory of relativity (GR) is verified by many observations. Nevertheless, some observations
seems not to fit GR and observed matter. As soon as 1933 Fritz Zwicky realized that the velocities of the Galaxies within  the Comma Cluster are way larger  than those predicted by the virial theorem in Newtonian theory \citet{zwicky}.  He remarked that the amount of matter needed to
  account for the velocities could be 400 times that of the visible matter. Which led to postulating  an unseen form of
   matter permeating the cluster.  Volders in 1959 remarked  that  stars in the periphery of the neighbor  spiral galaxy M33
   do not move as expected \citet{volders}.  The virial theorem in Newtonian Gravity  predicts that $MG/r \sim M v^2$, that is to say,
   the rotation curve should increase and at some point bend down  and the velocity should drop off as $1/\sqrt{r}$.
   In the seventies Rubin and Ford \citep{rubin1,rubin2} showed for a very large sample of spiral galaxies that this behavior
   is a general feature: velocities at the periphery of the galaxies do not bend down, attain a plateau at some velocity for each galaxy.
    In figure \ref{M33rc} we see a  rotation curves for the M33 galaxy describing this situation. In this paper we will attempt to show that such effects can be deduced from GR if retardation effects are not neglected \cite{YaRe1,ge,YaRe2,YaRe3}.
It should be stressed that the current approach does not require that velocities, $v$ are high, in fact the vast majority of galactic bodies (stars, gas) are substantially subluminal. In other words, the ratio of $\frac{v}{c} \ll 1$. Typical velocities in galaxies are ~100 km/s, which makes this ratio $0.001$ or smaller.
However, one should consider the fact that every gravitational system even if it is made of subluminal bodies has a retardation distance, beyond which the retardation effect cannot be neglected.  Every natural system such as stars and galaxies and even galactic clusters  exchange mass with its environment. For example, the sun losses mass through the solar wind and galaxies accrete gas from the intergalactic medium. This means that all natural gravitational systems have a finite retardation distance. The question is thus quantitative, how large is the retardation distance? The change of mass of the sun is quite small and thus the retardation distance of the solar system is quite large allowing us to neglect retardation effects within the solar system. However, for the M33 galaxy the velocity curve indicate that the retardation effects cannot be neglected beyond a certain distance which is calculated in section \ref{MOND} to be roughly 12,000 light years, similar analysis for other galaxies of different types has shown similar results \cite{Wagman}. We demonstrate in section \ref{dynmodel} using a detailed model that this does not require high velocity of gas or stars in or out of the galaxy and is perfectly consistent with the current observational knowledge of galactic and extra galactic material content and dynamics.

\section {Linear Approximation of GR}

Except for the extreme cases of compact objects (black holes and neutron stars) and the very early universe (big bang) one need not consider
the full non-linear Einstein equation. In most other cases of astronomical interest (galactic dynamics included) one can linearize those equations around the flat Lorentz metric $\eta_{\mn}$ such that\footnote{Private communication with the late Professor Donald Lynden-Bell}:
 \beq
 g_{\mn} = \eta_{\mn} + h_{\mn}, \quad \eta_{\mn} \equiv \ {\rm diag } \ (1,-1,-1,-1), \quad |h_{\mn}|\ll 1
 \label{lg}
 \enq
 One then defines the quantity:
 \beq
 \bar h_\mn \equiv h_\mn -  \frac{1}{2} \eta_\mn h, \quad h = \eta^{\mn} h_{\mn},
 \label{bh}
 \enq
 $\bar h_\mn = h_\mn $ for non diagonal terms. For diagonal terms:
 \beq
 \bar h = - h \Rightarrow  h_\mn = \bar h_\mn -  \frac{1}{2} \eta_\mn \bar h .
 \label{bh2}
 \enq
  It can be shown (\citet{Narlikar} page 75 exercise 37, see also \citep{Edd,Weinberg,MTW}), that one can  choose a gauge such that the Einstein equations are:
 \beq
\bh_{\mn, \alpha}{}^{\alpha}=-\frac{16 \pi G}{c^4} T_\mn , \qquad \bh_{\mu \alpha,}{}^{\alpha}=0.
\label{lineq1}
\enq
\Ern{lineq1} can always be integrated to take the form \citet{Jackson}\footnote{For reasons why the symmetry between space and time
is broken see \citep{Yahalom,Yahalomb}} (equation (6.47) given in page 245):
 \ber
& & \bh_{\mn}(\vec x, t) = -\frac{4 G}{c^4} \int \frac{T_\mn (\vec x', t-\frac{R}{c})}{R} d^3 x',
\nonumber \\
 t &\equiv& \frac{x^0}{c}, \quad \vec x \equiv x^a \quad a,b \in [1,2,3], \quad \vec R \equiv \vec x - \vec x', \quad R= |\vec R |.
\label{bhint}
\enr
The factor before the integral is small: $\frac{4 G}{c^4} \simeq 3.3 \ 10^{-44}$ hence in the above calculation
one can take $T_\mn$ which is zero order in $h_\ab$.
Let us now calculate the affine connection in the linear approximation:
\beq
\Gamma^\alpha_\mn = \frac{1}{2} \eta^\ab \left(h_{\beta \mu, \nu} + h_{\beta \nu, \mu} - h_{\mn, \beta}\right).
\label{affinel}
\enq
The affine connection has only first order terms, hence for a first order approximation of
$\Gamma^\alpha_\mn u^\mu u^\nu$ appearing in the geodesic,  $u^\mu u^\nu$ is zeroth order. In the zeroth order:
\beq
u^0=\frac{1}{\sqrt{1-\frac{v^2}{c^2}}},  \qquad u^a = \vec u =\frac{\frac{\vec v}{c}}{\sqrt{1-\frac{v^2}{c^2}}} , \qquad
\vec v \equiv  \frac{d \vec x}{d t}, \quad v= |\vec v|.
\label{uz}
\enq
For non relativistic velocities:
\beq
u^0 \simeq 1,  \qquad \vec u \simeq \frac{\vec v}{c} , \qquad u^a \ll u^0   \qquad {\rm for} \quad v \ll c.
\label{uzslo}
\enq
Inserting \ern{affinel} and \ern{uzslo} in the geodesic equation we arrive at the approximate form:
\beq
\frac{d v^a}{dt}\simeq - c^2 \Gamma^a_{00} = - c^2 \left( h^a_{0,0} - \frac{1}{2} h_{00,}{}^a \right)
\label{geol}
\enq
Let us now look at $T_\mn = (p+\rho c^2) u_\mu  u_\nu - p \ g_\mn$. In the current case $\rho c^2 \gg p$, combining this with  \ern{uzslo} we arrive at $T_{00} = \rho c^2 $ while all other components of the tensor
$T_\mn$ are significantly smaller. This implies that $\bar h_{00}$ is significantly larger than other components of the tensor
$\bar h_\mn$. Of course one should be careful and not deduce from the different magnitudes of quantities that such a difference
exist between their derivatives. In fact by the gauge condition in \ern{lineq1}:
\beq
\bar h_{\alpha 0,}{}^0=-\bar h_{\alpha a,}{}^a \qquad \Rightarrow
\bar h_{00,}{}^0=-\bar h_{0 a,}{}^a, \quad \bar h_{b0,}{}^0=-\bar h_{b a,}{}^a.
\label{gaugeim}
\enq
Hence the zeroth derivative of $\bar h_{00}$ (contains a $\frac{1}{c}$ factor) is the same order as the spatial derivative
of $\bar h_{0a}$ and like wise the zeroth derivative of $\bar h_{0a}$ (which appears implicitly in \ern{geol}) is the same order
of  the spatial derivative of $\bar h_{ab}$. However, it is safe to compare spatial derivatives of $\bar h_{00}$ and $\bar h_{ab}$
and conclude that the former is significantly larger than the later. Using \ern{bh2} and taking the above consideration into account
we write \ern{geol} as:
\beq
\frac{d v^a}{dt}\simeq \frac{c^2}{4} \bar h_{00,}{}^a \Rightarrow \frac{d \vec v}{dt} = - \vec \nabla \phi = \vec F,
\qquad \phi \equiv \frac{c^2}{4} \bar h_{00}
\label{geol2}
\enq
Thus $\phi$ is a gravitational potential of the motion which can be calculated using \ern{bhint}:
\ber
\phi &=& \frac{c^2}{4} \bar h_{00}
= -\frac{ G}{c^2} \int \frac{T_{00} (\vec x', t-\frac{R}{c})}{R} d^3 x'
\nonumber \\
& = & -G \int \frac{\rho (\vec x', t-\frac{R}{c})}{R} d^3 x'
\label{phi}
\enr
and $\vec F$ is the force per unit mass. If $\rho$ is static we are in the realm of the Newtonian instantaneous action at a distance theory.
However, it is unlikely that $\rho$ is static as a galaxy will attract mass from the intergalactic medium.

\section {Beyond the Newtonian Approximation}

The retardation time $\frac{R}{c}$ which may be a few tens of thousands of years is short with respect
to the time that the galactic density changes significantly. This means that we can write a Taylor series for the density:
\beq
\rho (\vec x', t-\frac{R}{c})=\sum_{n=0}^{\infty} \frac{1}{n!} \rho^{(n)} (\vec x', t) (-\frac{R}{c})^n,
\qquad \rho^{(n)}\equiv \frac{\partial^n \rho}{\partial t^n}.
\label{rhotay}
\enq
As in all expansions the
above equation is valid only for a certain environment of $t$ on the time axis which depends on the function $\rho (t)$, this environment shall be defined using the convergence radius $T_{max ~ \infty}$. That is \ern{rhotay} is valid only in the domain $[t-T_{max ~ \infty},t+T_{max ~ \infty}]$.
Inserting \ern{rhotay} into \ern{phi} and keeping the first three terms we will obtain:
\ber
\phi &=& -G \int \frac{\rho (\vec x', t)}{R} d^3 x' +  \frac{G}{c}\int \rho^{(1)} (\vec x', t) d^3 x'
\nonumber \\
&-& \frac{G}{2 c^2}\int R \rho^{(2)} (\vec x', t) d^3 x'
\label{phir}
\enr
As only second order terms are kept the above equation is a suitable approximation only for a smaller environment of $t$ on the time axis, this environment shall be defined using the time $T_{max ~ 2}<T_{max ~ \infty}$. That is \ern{phir} is valid only in the domain $[t-T_{max ~ 2},t+T_{max ~ 2}]$. Or for distances satisfying:
\beq
R < c \ T_{max~ 2} \equiv R_{max}
\label{Rmax}
\enq
This means that current expansion is related to the near field case, this is acceptable since the extension of the rotation curve in galaxies
is the same order of magnitude as the size of the galaxy itself. An opposite case in which the size of the object is much smaller than the distance to the observer will result in a different approximation to \ern{bhint} leading to the famous quadruple equation of gravitational radiation as predicted by Einstein \cite{Einstein2} and verified indirectly in 1993 by Russell A. Hulse and Joseph H. Taylor, Jr.  for which they received the Nobel Prize in Physics. The discovery and observation of the Hulse-Taylor binary pulsar offered the first indirect evidence of the existence of gravitational waves \cite{Taylor}. On 11 February 2016, the LIGO and Virgo Scientific Collaboration announced they had made the first direct observation of gravitational waves. The observation was made five months earlier, on 14 September 2015, using the Advanced LIGO detectors. The gravitational waves originated from the merging of a binary black hole system \cite{Castelvecchi}. Thus the current paper involve a near field application of gravitational radiation while previous art discusses far field results.

The first term will provide the Newtonian potential, the second term does not contribute, the third term will result in the lower order correction to the Newtonian theory:
\beq
 \phi_r = - \frac{G}{2 c^2} \int  R \rho^{(2)} (\vec x', t) d^3 x'
\label{phir2}
\enq
The total force per unit mass:
\ber
\vec F &=& \vec F_N + \vec F_r
\nonumber \\
 \vec F_N &=& - \vec \nabla \phi_N =  -G \int \frac{\rho (\vec x',t)}{R^2} \hat R d^3 x', \qquad \hat R \equiv \frac{\vec R}{R}
\nonumber \\
 \vec F_r &\equiv& - \vec \nabla \phi_r =  \frac{G}{2 c^2} \int  \rho^{(2)} (\vec x', t) \hat R d^3 x'
\label{Fr}
\enr
While the Newtonian force $\vec F_N$ is always attractive the retardation force $\vec F_r$ can be
either attractive or repulsive. Also notice that while the Newtonian force decreases as $\frac{1}{R^2}$ , the retardation force is independent of distance as long as the Taylor approximation of \ern{rhotay} is valid. For short
distances the Newtonian force is dominant but as the distances increase the retardation force becomes dominant. Newtonian force can be
neglected for distances significantly larger than the retardation distance:
\beq
 R \gg R_r \equiv c t_r
\label{Rr}
\enq
$t_r$ is the typical duration in which the density $\rho$ changes. Of course for $R\ll R_r$ the retardation effect can be neglected and only
Newtonian forces should be considered. For large distances $r=|\vec x|\rightarrow\infty$ such that
 $\hat R \simeq \frac{\vec x}{|\vec x|} \equiv \hat r$ we obtain:
\beq
\vec F_r =  \frac{G}{2 c^2} \hat r \int  \rho^{(2)} (\vec x', t)  d^3 x' =  \frac{G}{2 c^2} \hat r \ddot{M}, \qquad
\ddot{M} \equiv \frac{d^2 M}{dt^2}.
\label{Fr2}
\enq
Now as the galaxy attracts intergalactic gas its mass increases thus $\dot{M}>0$,
however, as the intergalactic gas is depleted the rate at which the mass increases must decrease hence $\ddot{M}<0$. Thus in the
galactic case:
\beq
\vec F_r =  - \frac{G}{2 c^2}  |\ddot{M}| \hat r
\label{Fr3}
\enq
and the retardation force is attractive.

\section{Dark Matter}

In what circumstances can one confuse retardation with the effect of a non existent "dark matter"? Let us ignore retardation effects and suppose that radial velocities are a result of some mysterious dark matter.
In this case we can write for a spherically symmetric mass distribution \citet{Binney}:
\beq
-\frac{v_c^2}{r} \hat r = \vec F_d = - \frac{G M_d(r)}{r^2} \hat r
\label{Fd}
\enq
$v_c$ is the speed of a test particle of constant radius $r$ and $M_d(r)$ is the amount of dark matter inside the radius $r$.
Comparing \ern{Fd} and \ern{Fr3} we see that the "dark matter" mass can be calculated as follows:
\beq
M_d(r) =   \frac{r^2 |\ddot{M}|}{2  c^2}
\label{Fr4}
\enq
Now since:
\beq
M_d(r) =  4 \pi \int_{0}^{r} r'^2 \rho_d (r') dr', \qquad \frac{d M_d(r) }{dr} = 4 \pi r^2 \rho_d (r)
\label{Md}
\enq
it follows:
\beq
\rho_d (r) = \frac{|\ddot{M}|}{4 \pi c^2 r}.
\label{rhod}
\enq
This is consistent with observational data of \citet{Corbelli} who concluded that the "dark matter" density
decreases as $r^{-1.3}$ for M33, however, notice that \citet{Corbelli} does not consider asymptotic distances as the rotation curve of M33 is given up to a distance 50,000 light years where the radius of M33 is about 30,000 light years, hardly asymptotic.
The equation:
\beq
-\frac{v_c^2}{r} \hat r = \vec F_r = - \frac{G}{2 c^2}  |\ddot{M}| \hat r
\label{vcra}
\enq
results in the asymptotic value for the velocity which is valid for distances much large than the galactic radius and the retardation distance (as otherwise one needs to take into account the Newtonian contribution as well):
\beq
v_c = \sqrt{\frac{G}{2 c^2}  |\ddot{M}| r}.
\label{vc}
\enq
For distances smaller than this, one needs to calculate both the retardation and Newtonian contributions
as described in \ern{Fr} this will be done below.

\section{MOND}
\label{MOND}

Another approach to explaining galactic rotation curves is the claim that either the laws of dynamics (Newton's second law) or
the laws of Gravitation (GR) should be modified. This approach championed by Milgrom is denoted "MOND" (Modified Newtonian dynamics) \citet{Mond}. In one version of this approach Newton's law  of gravity
is modified:
\beq
 \vec F_M = - \frac{G M}{\mu(\frac{a}{a_0}) r^2} \hat r
\label{FM}
\enq
In the above $\mu$ is the interpolation function that should be $1$ for $a_0\ll a$. Let us assume:
\beq
  \mu(x) = \frac{x}{\sqrt{1+x^2}} \quad \Rightarrow \quad  \mu(\frac{a}{a_0})=\frac{1}{\sqrt{1+(\frac{a_0}{a})^2}}
\label{mu}
\enq
If $a_0\gg a$, $\mu \simeq \frac{a}{a_0}$. A test particle
revolving in a constant radius will have centrifugal acceleration $a=\frac{v^2}{r}$ and thus:
\beq
 \vec F_M = - \frac{G M a_0}{v^2 \ r} \hat r
\label{FM2}
\enq
For $v$ constant at a far away distance this expression is similar to the retardation force
and thus:
 \beq
 |\ddot{M}| = \frac{2 M a_0 c^2}{v^2 \ r}.
\label{ddM}
\enq
Milgrom found $a_0 = 1.2 \ 10^{-10} m s^{-2}$ to be most fitting to the data. The baryonic mass of the M33 galaxy is $2 \ 10^{40} kg$
\cite{Corbelli2} and the velocity at $15.33 \ kpc$ from the center of the galaxy is $135,640 \ m s^{-1}$. We thus obtain $|\ddot{M}| \simeq  4.94 \ 10^{16}  kg s^{-2}$ and a retardation time of:
\beq
t_r\equiv \sqrt{\frac{M}{|\ddot{M}|}} \simeq 6.35 \ 10^{11} \ s
\label{deltat}
\enq
 This amounts to a typical accumulation acceleration time scale of $t_r\simeq 20,129$ years and retardation distance of:
\beq
R_r = c t_r\simeq 20,129 {\rm \ light \ years}
\label{Rr2}
\enq
which seems reasonable according to figure \ref{M33rc}. We under line that this is based only on the {\bf current} estimation of the second derivative of mass, we make no claim  on the past or future values of $\ddot{M}$ nor is there any claim in this paper on the value of
$\dot{M}$ at any time or the value of $M$ in the past or the future. It is obvious that such questions involve a full theory of galaxy formation
taking into account retardation forces, such a theory is beyond the scope of this paper and will need to wait for a future paper. It is also clear that the current value of $\ddot{M}$ is unsustainable over the entire history of the galaxy. To see this let us assume that $\ddot{M}$ is constant. This leads to:
\beq
\dot{M}(t)= \dot{M}(0) + t \ddot{M} = \dot{M}(0) - t |\ddot{M}|
\label{dMt}
\enq
Let assume that there is no significant mass accumulation at the current time $T$, that is $\dot{M}(T)=0$. In this case:
\beq
 \dot{M}(0) = T |\ddot{M}|
\label{dMt0}
\enq
Assuming that the age of the galaxy is about the age of the universe:  $T \simeq 13 \ 10^9 \ years = 4.1 \ 10^{17} \ s$ .
Hence the initial mass accumulation rate is about $\dot{M}(0) \simeq 2.12 \ 10^{34} \ kg \ s^{-1}$. Now integrating \ern{dMt} we arrive
at an equation for the mass of galaxy at any time:
\beq
M(t) = M(0) + \dot{M}(0) t + \frac{1}{2} \ddot{M} t^2
\label{Mt2}
\enq
Assuming that initially there was no significant amount of mass: $M(0)=0$ and taking into account \ern{dMt0} we arrive at:
\beq
M(t) =\ddot{M} t (\frac{1}{2} t - T) \Rightarrow  M(T) =-\frac{1}{2} \ddot{M} T^2= \frac{1}{2} |\ddot{M}| T^2
\label{Mt3}
\enq
Plugging in the mass accumulation decrease rate we arrive at $ M(T) \simeq 4.54 \ 10^{51}$ which is clearly 11 orders of magnitude greater
than the known mass of the galaxy. On the other hand if we assume that $M(T)=M$ the current mass of the galaxy this will lead
to $|\ddot{M}| \simeq  2.37 \ 10^{5} \  kg \ s^{-2}$ which is 11 orders of magnitude less than what is needed to explain the current galactic
rotation curves. This indicates that $|\ddot{M}| $ has increased considerably over time due to the depletion of the surrounding gas and that
retardation forces are less significant in the early stages of galactic formation. Indeed in young galaxies the effect of retardation seems insignificant \citet{Dokkum}, while for older galaxies it seems like
somebody is pressing hard on the brakes of mass accumulation. The change of many order of magnitudes in $|\ddot{M}|$ suggest exponential growth,
hence the following model is suggested:
\beq
M(t) = M(0)+ \left(\dot{M}(0)-\tau \ddot{M}(0)\right) t + \tau^2 \ddot{M}(0) \left(e^{\frac{t}{\tau}} - 1\right), \quad \tau>0
\label{Mt4}
\enq
\beq
\dot{M}(t) = \dot{M}(0) + \tau \ddot{M}(0) \left(e^{\frac{t}{\tau}} - 1\right)
= \dot{M}(0) -\tau |\ddot{M}(0)| \left(e^{\frac{t}{\tau}} - 1\right)
\label{dMt4}
\enq
\beq
\ddot{M}(t) = \ddot{M}(0) e^{\frac{t}{\tau}}.
\label{ddMt4}
\enq
If we assume $\dot{M}(t) > 0$ it follows from \ern{dMt4} that there is a maximal duration for the above
model to be valid:
\beq
t_{max1} = \tau \ln \left(\frac{\dot{M}(0)}{\tau |\ddot{M}(0)|} + 1 \right).
\label{dMt5}
\enq
This is the time it takes the intergalactic gas to deplete (see section \ref{dynmodel}) and thus mass accumulation stops.
We shall now partition the mass into a linear and exponential growing part as follows:
\beq
M(t) =  M_l (t) + M_e (t) = M_l (t) - |M_e (t)| > 0 \Rightarrow M_l (t) > |M_e (t)|.
\label{Mt5a}
\enq
\ber
M_l (t) & = &  M(0) + \tau^2 |\ddot{M}(0)| + \left(\dot{M}(0)+ \tau |\ddot{M}(0)|\right) t > 0,
\nonumber \\
M_e(t) & = &  \tau^2 \ddot{M}(t) =  - \tau^2 |\ddot{M}(0)| e^{\frac{t}{\tau}} < 0.
\label{Mt6}
\enr
Dividing \ern{Mt5a} by $|\ddot{M}(t)|$  and using the definition of \ern{deltat} we have:
\beq
t_r^2  = \frac{M_l (t)}{|\ddot{M}(t)|} - \tau^2.
\label{Mt7}
\enq
We define the slow retardation time:
\beq
t_{rs} (t) \equiv  \sqrt{\frac{M_l (t)}{|\ddot{M}(0)|}},
\label{Mt8}
\enq
the square of this retardation time is linearly dependent on time. Hence:
\beq
t_r^2  =t_{rs}^2 e^{-\frac{t}{\tau}} - \tau^2
\label{Mt9}
\enq
It is obvious that $t_{rs} > \tau$ and it is also obvious that for any $t<t_{max1}$ then $t_{rs} e^{-\frac{t}{2\tau}} > \tau$ because
the total mass is not reduced but only  mass accumulation slows down (in other words we assume that the galaxy does not eject matter thus reducing its mass). \Ern{Mt9} shows that there is no simple relation between the retardation time of the galaxy and the typical time $\tau$ associated with the second derivative of mass.

We conclude this section by deriving constraints from the assumption that apparent luminosity and rotation curve have not changed in a detectable way since observations of the M33 galaxy have begun. If the mass to light ratio is assumed constant this means that the mass of the galaxy has not changed significantly during the said period. Furthermore, the retardation time  which has a significant effect on the galactic rotation curve is a square root of the ratio of the galactic mass by the second derivative of the same mass. This means that if the retardation time (and retardation distance) have not changed considerably during the duration of observations then the second derivative must also be approximately constant during that period. The M33 Triangulum Galaxy was probably discovered by the Italian astronomer Giovanni Battista Hodierna before 1654. In his work De systemate orbis cometici; deque admirandis coeli caracteribus ("About the systematics of the cometary orbit, and about the admirable objects of the sky"), he listed it as a cloud-like nebulosity or obscuration and gave the cryptic description, "near the Triangle hinc inde". This is in reference to the constellation of Triangulum as a pair of triangles. The magnitude of the object matches M33, so it is most likely a reference to the Triangulum galaxy \cite{Fodera}. This observation took place 366 years ago, however, there was no measurement of apparent or real luminosity of M33 and of course no measurement of doppler shifts were available at that time. It was only Hubble that established the distance to this object and thus declared it an independent galaxy far away from the milky way \cite{Van}, thus enabling the calculation of the real luminosity and mass of M33. Doppler shifts of spectral lines were obtained even later thus enabling the deduction of rotation curves. In what follows we shall assume an observation duration of:
\beq
\delta t = {\rm 400 \ years}
\label{Mt10}
\enq
As $\ddot{M}(t)$ is approximately constant during the said duration, it follows that:
\beq
|\frac{d^3 M(t)}{dt^3}\delta t| \ll |\ddot{M}(t)| .
\label{ddMt11}
\enq
Taking into account \ern{ddMt4} this gives a lower bound to $\tau$:
\beq
\delta t \ll \tau.
\label{ddMt12}
\enq
This is not a strong bound as $\tau$ as it can be anything from 100,000 years to ten times the age of the universe or even larger. Similarly
the fact that mass of the galaxy has not changed considerably during the observation duration indicates that:
 \beq
|\dot{M}(t)| \delta t| \ll |M(t)|  \Rightarrow  \dot{M}(t) \ll \dot{M}_{max} \equiv \frac{M(t)}{\delta t}
\label{Mt11}
\enq
The upper bound on the mass derivative is thus about $\dot{M}_{max} \simeq 1.6 \ 10^{30} $ kg/s, in which we have taken into account baryonic mass of the M33 galaxy to be about $2 \ 10^{40}$ kg \cite{Corbelli2}. Using \ern{dMt4} we arrive with an upper bound to $\dot{M}(0)$:
\beq
\dot{M}(0) \ll \dot{M}_{max} + \tau  \left(|\ddot{M}(t)| - |\ddot{M}(0)| \right).
\label{dMt4b}
\enq
Notice, however, that without a handle on $\tau$ and $|\ddot{M}(0)|$ this does not yield any definite information.
In the next section we turn our attention to the dynamical processes that are responsible for the second derivative of the galactic mass and show
how additional information can be obtained from observations.

\section{A Dynamical Model}
\label{dynmodel}

\subsection{Basic equations}

The mass accumulation model described in the previous section is based on a fitting of the second derivative of the galactic mass to the
galactic rotation curve. It is intuitively obvious that as mass is accumulated in the galaxy it must be depleted in the intergalactic medium.
This is due to the fact that the total mass is conserved still it is of interest to see if this intuition is compatible with a model of gas dynamics. For simplicity we assume that the gas is a barotropic ideal fluid and its dynamics is described by the Euler and continuity equations as follows:
\beq
\frac{\partial{\rho}}{\partial t} + \vec \nabla \cdot (\rho \vec v ) = 0
\label{masscon}
\enq
\beq
\frac{d \vec v}{d t} \equiv
\frac{\partial \vec v}{\partial t}+(\vec v \cdot \vec \nabla)\vec v  = -\frac{\vec \nabla p (\rho)}{\rho} - \vec \nabla \phi
\label{Euler}
\enq
In which the pressure $p (\rho)$ is assumed to be a given function of the density,  $\frac{\partial }{\partial t}$
is a partial temporal derivative, $\vec \nabla$ has its standard meaning in vector analysis and $\frac{d }{d t}$ is the material temporal derivative. We have neglected viscosity terms due to the gas low density.

\subsection{General considerations}

Let us now take a partial temporal derivative of \ern{masscon} leading to:
\beq
\frac{\partial^2 {\rho}}{\partial t^2} + \vec \nabla \cdot (\frac{\partial{\rho}}{\partial t} \vec v +
 \rho \frac{\partial \vec v}{\partial t} ) = 0.
\label{masscon2b}
\enq
Using \ern{masscon} again we  obtain the expression:
\beq
\frac{\partial^2 {\rho}}{\partial t^2} = \vec \nabla \cdot \left(\vec \nabla \cdot (\rho \vec v ) \vec v  -
 \rho \frac{\partial \vec v}{\partial t} \right) .
\label{masscon3}
\enq
We divide the left and right hand sides of the equation by $c^2$ as in \ern{Fr} and obtain:
\beq
\frac{1}{c^2}\frac{\partial^2 {\rho}}{\partial t^2} = \vec \nabla \cdot \left( \frac{\vec v}{c} \left(\frac{\vec v}{c} \cdot \vec \nabla \rho +
\rho \vec \nabla \cdot (\frac{\vec v}{c} )\right)   -
 \rho \frac{1}{c} \frac{\partial \frac{\vec v}{c}}{\partial t} \right) .
\label{masscon4}
\enq
Since $\frac{\vec v}{c} $ is rather small in galaxies it follows that $\frac{1}{c^2}\frac{\partial^2 {\rho}}{\partial t^2}$ is also small unless the density or the velocity have significant spatial derivatives. A significant acceleration $\frac{\partial \frac{\vec v}{c}}{\partial t}$ resulting from a considerable force can also have a decisive effect. The depletion of intergalactic gas can indeed cause such gradients as we describe below using a detailed model. Finally taking the volume integral of the left and right hand sides of \ern{masscon4} and using Gauss theorem we arrive at the following equation:
\beq
\frac{1}{c^2}\ddot{M}=\frac{1}{c^2} \int \frac{\partial^2 {\rho}}{\partial t^2} d^3 x = \oint d \vec S \cdot \left( \frac{\vec v}{c} \left(\frac{\vec v}{c} \cdot \vec \nabla \rho + \rho \vec \nabla \cdot (\frac{\vec v}{c} )\right)   -
 \rho \frac{1}{c} \frac{\partial \frac{\vec v}{c}}{\partial t} \right) .
\label{masscon5}
\enq
The surface integral is taken over a surface encapsulating the galaxy.

\subsection{Detailed Model}

\begin{figure}
\includegraphics[width=0.9\columnwidth]{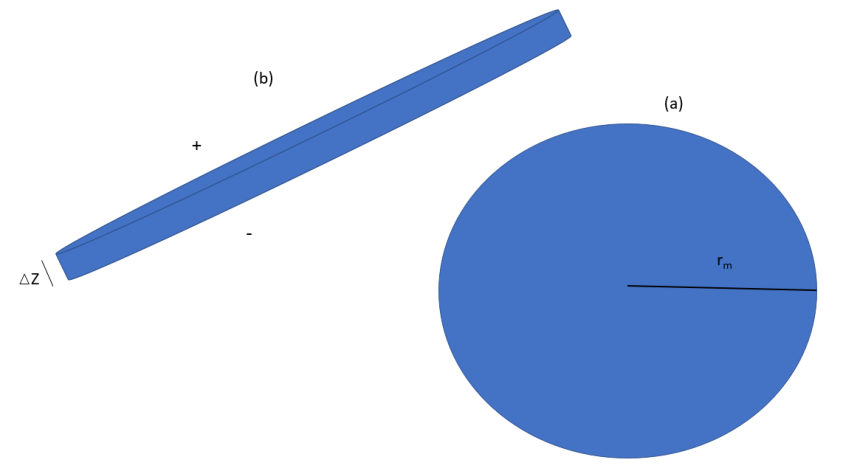}
 \caption{An idealized cylindrical galaxy from different perspectives. (a) from above (b) tilted edge perspective.}
 \label{idgalax}
\end{figure}

We shall now assume that the system is described by cylindrical coordinates $\br,\theta,z$ in which $z=0$ is the galactic plane. For simplicity we assume axial symmetry, hence all variables are independent of $\theta$. Moreover, the mass influx coming from above and below the galaxy is much more significant as compared to the influx coming from the galactic edge. This is due to the large difference of the galaxy surfaces perpendicular to the z axis compared to the area of its edge. The area of the surface of the galaxy which is perpendicular to the $z$ axis is:
\beq
S_z= S_{z+}+S_{z-} = \pi r_m^2 + \pi r_m^2 = 2 \pi r_m^2
\label{galsz}
\enq
in which $S_z$ is the total surface area of the galaxy perpendicular to the $z$ axis, $S_{z+}$ is the upper area of the surface of the galaxy perpendicular to the $z$ axis, $S_{z-}$ is the lower area of the surface of galaxy perpendicular to the $z$ axis and $r_m$ is the galactic radius
(see figure \ref{idgalax}).

\begin{figure}
\includegraphics[width=0.9\columnwidth]{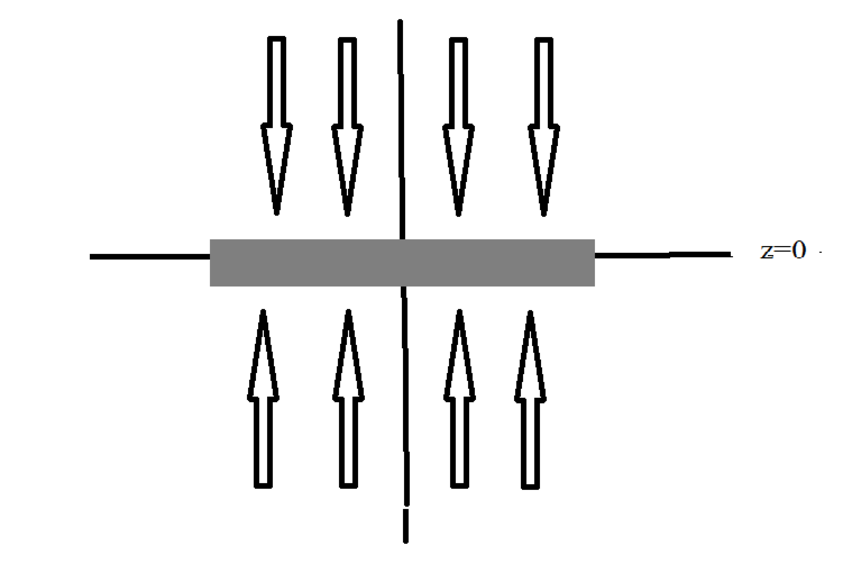}
 \caption{A schematic view of the galactic influx from a side view.}
 \label{influx}
\end{figure}

\begin{figure}
\includegraphics[width=0.9\columnwidth]{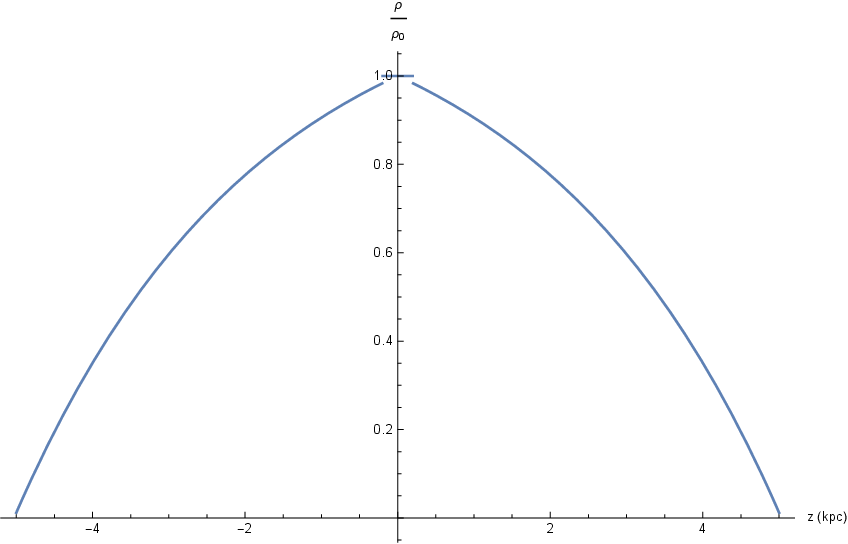}
 \caption{An initial density profile out side the galactic plane. In which $\rho_0 = \rho_1+\rho_2, \frac{\rho_2}{\rho_1}=-0.2$ and $z_i=5$ (kpc) and $k=0.32 \ \rm (kpc^{-1})$.}
 \label{indenprof}
\end{figure}

\begin{figure}
\includegraphics[width=0.9\columnwidth]{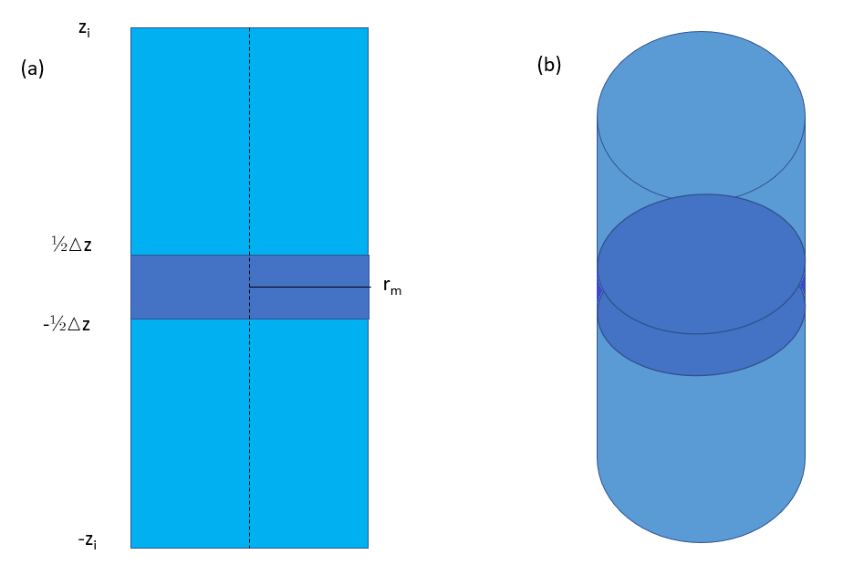}
 \caption{The mass column defined by the galaxy (a) Side view (b) Three dimensional view.}
 \label{masscol}
\end{figure}

The area of the surface of the galactic edge  with thickness $\Delta z$ is:
\beq
S_e= 2 \pi r_m \Delta z.
\label{galse}
\enq
And thus the ratio of the surfaces area is:
\beq
\frac{S_e}{S_z}= \frac{\Delta z}{r_m}.
\label{galsra}
\enq
Typical values of $\Delta z$ is about $0.4$ kilo parsec and $r_m$ is about $17$ kilo parsec (for M33) giving an area ratio of about $1\%$.
In such circumstances the edge mass influx is less important and we  can assume a velocity field of the form:
\beq
\vec v = v_z (\br,z,t) \hat z + v_\theta (\br,z,t) \hat \theta.
\label{velfield}
\enq
$\hat z$ and $\hat \theta$ are unit vectors in the $z$ and $\theta$ directions respectively.
The influx is described schematically in figure \ref{influx}.
In this case the continuity \ern{masscon} will take the form:
\beq
\frac{\partial{\rho}}{\partial t} + \frac{\partial{(\rho v_z)}}{\partial z} = 0
\label{masscon2}
\enq
Defining the quantity:
\beq
\gamma \equiv \rho v_z \Rightarrow  \rho = \frac{\gamma}{v_z}
\label{gamma}
\enq
and using the above definition \ern{masscon2} takes the form:
\beq
\frac{\partial{(\frac{\gamma}{v_z})}}{\partial t} + \frac{\partial{\gamma}}{\partial z} = 0
\label{gammaeq}
\enq
Assuming for simplicity that $v_z$ is stationary and defining the auxiliary variable $t_z$:
\beq
t_z \equiv \int \frac{dz}{v_z}
\label{tz}
\enq
we arrive at the equations:
\beq
\frac{\partial{\gamma}}{\partial t} + \frac{\partial{\gamma}}{\partial t_z} = 0.
\label{gammaeq2}
\enq
This equation can be solved easily as follows:
\beq
\gamma (\br,z,t) = f(t-t_z), \qquad  f(-t_z) =  \gamma (\br,z,0) = v_z \rho (\br,z,0)
\label{gamma2}
\enq
for the function $f(x)$ which is fixed by the density initial conditions and the velocity profile. Let us now turn our attention to the Euler \ern{Euler}, for stationary flows it takes the form:
\beq
(\vec v \cdot \vec \nabla)\vec v  = -\frac{\vec \nabla p (\rho)}{\rho} - \vec \nabla \phi
\label{Eulers}
\enq
According to \ern{velfield} :
\beq
\vec v \cdot \vec \nabla = v_z \frac{\partial}{\partial z} +\frac{v_\theta}{\br} \frac{\partial}{\partial \theta}
\label{vcon}
\enq
Now writing \ern{Eulers} in terms of its components we arrive at the following equations:
\beq
v_z \frac{\partial v_z }{\partial z}  = -\frac{1}{\rho} \frac{\partial p }{\partial z} - \frac{\partial \phi }{\partial z}
\label{Eulersz}
\enq
\beq
-  \frac{v_\theta^2}{\br}   = -\frac{1}{\rho} \frac{\partial p }{\partial \br} - \frac{\partial \phi }{\partial \br},
 \qquad \left(\frac{\partial \hat{\theta} }{\partial \theta} = - \hat{\br}\right).
\label{Eulersr}
\enq
It is usually assumed that the radial pressure gradients are negligible with respect to the gravitational forces and thus we arrive at the
equation:
\beq
 \frac{v_\theta^2}{\br} \simeq  \frac{\partial \phi }{\partial \br},
\label{Eulersr2}
\enq
As for the $z$ component equation it can be easily written in terms of the specific enthalpy $w(\rho) = \int \frac{dP}{\rho}$ in the form:
\beq
\frac{\partial }{\partial z}  \left(\frac{1}{2} v_z^2 + w(\rho) + \phi \right) = 0 \Rightarrow
\frac{1}{2} v_z^2 + w(\rho) + \phi = C(r,t).
\label{Eulersz3}
\enq
We recall that $\rho$ depends on $v_z$ through \ern{gamma} and \ern{gamma2}:
\beq
\rho(r,z,t) =  \frac{\gamma}{v_z} = \frac{f(t- \int \frac{dz}{v_z})}{v_z}
\label{rho2}
\enq
As both the specific enthalpy and the gravitational potential are dependent on the density, \ern{Eulersz3} turns into a rather complicated
nonlinear integral equation for $v_z$. However, many galaxies are flattened structures, hence it can thus be assumed that the pressure $z$ gradients are significant as one approaches the galactic plane. We will thus assume for the sake of simplicity that the pressure gradients balance the gravitational pull of the galaxy and thus $v_z$ is just a function of $r$ in which case the convective derivative of $v_z$ vanishes. The above assumption holds below and above the galactic plane but not at the galactic plane itself. This suggests the following simple model for the velocity $v_z$ (see figure \ref{influx}):
\beq
v_z=\left\{
      \begin{array}{cc}
        -|v_z| & z>0 \\
        |v_z| & z<0 \\
      \end{array}
    \right.
\label{vz}
\enq
in which $|v_z|$ is a known function of $\br$. The velocity field is discontinuous at the galactic plane due to our simplification assumptions, but of course need not be so in reality. We also assume for simplicity that the velocity field $|v_z|$ is constant for $\br<r_m$ and  vanishes for $\br>r_m$. According to \ern{gamma2} the time dependent density profile is fixed by the density initial conditions. In this section we will deal with the density profile outside the galactic plane and will leave the discussion of the density profile in and near the galactic plane to the next section. We consider an initial density profile as follows:
\beq
\rho_o(\br,z,0)=re(z)\left[\rho_1 (\br) + \rho_2 (\br) e^{k|z|}\right],
 \qquad re(z)=\left\{           \begin{array}{cc}
                                                                                       1 & |z|<z_i \\
                                                                                       0 & |z| \ge z_i \\
                                                                                     \end{array}
                                                                                   \right.
\label{rho0}
\enq
in which the rectangular function $re(z)$ keeps the exponential function from diverging. The density profile is depicted in figure \ref{indenprof}.
We assume that $\rho_2$ is negative and thus the density becomes dilute at distances far from the galactic plane.
As $v_z$ is constant both above and below the galactic plane, $t_z=\frac{z}{v_z}$ up to a constant. And now it is easy to deduce from \ern{gamma2} the functional form of $f(\beta)$:
\beq
f(\beta) = v_z re (-v_z \beta) [\rho_1 + \rho_2 e^{k|v_z \beta|}]
\label{fder}
\enq
And hence according to \ern{rho3} the time dependent density function for matter outside the galactic plane is obtained:
\beq
\rho_o(\br,z,t) =  \frac{\gamma}{v_z} = \frac{f(t- \int \frac{dz}{v_z})}{v_z}=  re (z - v_z t) [\rho_1 (\br)  + \rho_2 (\br) e^{k|z - v_z t|}]
\label{rho3}
\enq
The density of matter outside the galactic plane will vanish for $t>t_m = \frac{z_i}{|v_z|}$, hence we will discuss only the duration of $t<t_m$.
Let us look at the mass contained in the cylinder defined by the galaxy  (see figure \ref{masscol}) and let us assume that the total mass in that cylinder is $M_T$. Now the
mass outside the galactic disk will be:
\beq
M_o (t)= 2\pi \left[\int_{-z_i}^{-\frac{1}{2} \Delta z} dz \int_{0}^{r_m} d \br \br \rho_o(\br,z,t) +
 \int_{\frac{1}{2} \Delta z}^{z_i} dz \int_{0}^{r_m} d \br \br \rho_o(\br,z,t) \right]
\label{Mo}
\enq
Hence the mass in the galactic disk is:
\beq
M(t) = M_T - M_o (t)
\label{Mg}
\enq
And the galactic mass derivatives are:
\beq
\dot{M}(t) =  - \dot{M_o}(t), \qquad \ddot{M}(t) =  - \ddot{M_o}(t)
\label{Mgd}
\enq
Inserting \ern{rho3} into \ern{Mo} we may calculate $M_o (t)$:
\beq
M_o (t)= 2 \left[\lambda_1 \left(z_i-|v_z|t-\hdz\right) + \frac{\lambda_2}{k} \left(e^{k z_i}- e^{k (|v_z|t +\hdz)}\right)\right]
\label{Mo2}
\enq
in which:
\beq
\lambda_1 \equiv 2 \pi \int_{0}^{r_m} d \br \br \rho_1 (\br), \qquad \lambda_2 \equiv 2 \pi \int_{0}^{r_m} d \br \br \rho_2 (\br).
\label{lam}
\enq
Now calculating the second derivative of $M_o (t)$ and using \ern{Mgd} leads to the result:
\beq
 \ddot{M}(t) =  - \ddot{M_o}(t) = 2 k |v_z|^2 e^{\hdz k} \lambda_2  e^{k |v_z|t}.
\label{Mgd2}
\enq
Comparing \ern{Mgd2} with \ern{ddMt4} leads to the following identification:
\beq
 \frac{1}{\tau} = k |v_z|, \qquad \ddot{M}(0) = 2 k |v_z|^2 e^{\hdz k} \lambda_2
\label{Mgd3}
\enq
which means that we must have $\lambda_2<0$ according to \ern{Fr3} in order to assure an attractive force.
As the velocities near the galactic plane can be determined using the method of Doppler shifts, and the density gradient
can be determined by looking at the gradient of the luminosity profile and also by looking at the gradient of the width of the spectral lines
 above the galactic plane, this allows a plausible observational approach to evaluate $\tau$.
Next we calculate $\dot{M}(t)$, using \ern{Mgd} and \ern{Mo2} we will obtain:
\beq
 \dot{M}(t) =  - \dot{M_o}(t) = 2 |v_z| \lambda_1 +  2  |v_z| e^{\hdz k} \lambda_2  e^{k |v_z|t}.
\label{Mgd4}
\enq
Hence:
\beq
 \dot{M}(0) =   2 |v_z| \lambda_1 +  2  |v_z| e^{\hdz k} \lambda_2  =  2 |v_z| \lambda_1 +\tau \ddot{M}(0)
\label{Mgd5}
\enq
Thus $\lambda_1$ is:
\beq
 \lambda_1 = \frac{1}{ 2 |v_z|} \left[\dot{M}(0)  - \tau \ddot{M}(0) \right]
\label{Mgd6}
\enq
Inserting \ern{Mgd6} and \ern{Mgd3} into \ern{Mgd4} leads back to \ern{dMt4}. Finally, combining equations (\ref{Mg},\ref{Mo2},\ref{Mgd3}, \ref{Mgd6}) and noticing that:
\beq
{M}(0) =  M_T  - \frac{z_i-\hdz}{  |v_z|} \left(\dot{M}(0)  - \tau \ddot{M}(0) \right) - \tau^2 \ddot M (0)
\left( e^{k(z_i -\hdz)} - 1 \right)
\label{Mgd7}
\enq
we arrive back at \ern{Mt4}. Hence the dynamical model presented in this section is compatible with the mass model of the previous one.

\section{Non Asymptotic Rotation Curves}

To calculate the non asymptotic rotation curve we will use \ern{Eulersr2}. The gravitational potential will be evaluated using \ern{phir}  which is composed of both the Newtonian $\phi_N$ and retardation $\phi_r$ parts such that:
\beq
 \frac{v_\theta^2}{\br} =  \frac{\partial \phi }{\partial \br} =\frac{\partial \phi_N }{\partial \br} + \frac{\partial \phi_r }{\partial \br} ,
\label{Eulersr3}
\enq
To do this we first need a density profile describing the mass distribution in and near the galactic plane.

\subsection{General Considerations}

We assume a density distribution of the form:
\beq
\rho(\vec x,t)=\rho_a (\vec x) + \rho_b (\vec x)g'(t)
 \label{rhog}
\enq
Although the density profiles $\rho_a$ and $\rho_b$ need not have similar forms we will assume for simplicity that they do. Hence
$\rho_a=\rho_b$ and defining $g(t) = 1 + g'(t)$  we obtain a density distribution of the form:
\beq
\rho(\vec x,t) = g(t) \rho_a (\vec x)
\label{rhotd}
\enq
Hence:
\beq
M(t) = \int d^3 x' \rho(\vec x',t) = g(t) \int d^3 x' \rho_a (\vec x')
\label{M}
\enq
And:
\beq
\ddot{M}(t) =  \ddot{g}(t) \int d^3 x' \rho_a (\vec x')
\label{ddM2}
\enq
Which leads to the result:
\beq
  |\ddot{g}(t)| = \frac{|\ddot{M}(t)|}{M(t)} g(t)  = \frac{c^2}{R_r^2} g(t)
\label{ddg}
\enq
Inserting \ern{rhotd} and \ern{ddg} into \ern{phir2} leads to the following form of the retardation potential:
\beq
 \phi_r = - \frac{G}{2 c^2} \ddot{g}(t) \int  R \rho_a (\vec x') d^3 x' =  \frac{G}{2 R_r^2} \int  R \rho (\vec x',t) d^3 x'.
\label{phir3}
\enq
Now we introduce the dimensionless quantities:
\beq
\tilde{\rho} \equiv \frac{\rho}{\rho_c}, \qquad \tilde{x} \equiv \frac{\vec x}{R_s}
\label{dless}
\enq
in terms of a typical density $\rho_c$ and a typical scale $R_s$. From now on we will consider all time dependent quantities as being given at the time the galaxy is being observed. This has nothing to do with the present situation in a certain galactic system that may be millions or billions year in the future with respect to the data available to us,  which of course is also the result of retardation. The dependence of the temporal variable $t$ will thus be omitted. Using \ern{dless}, \ern{M} will take the form:
\beq
M = \int d^3 x' \rho(\vec x') = \rho_c R_s^3 \int d^3 \tilde x' \tilde \rho(\tilde x')
\label{M2}
\enq
We now define the dimensionless constant:
\beq
\Lambda = \int d^3 \tilde x' \tilde \rho(\tilde x')
\label{lam2}
\enq
In terms of which:
\beq
M =  \Lambda \rho_c R_s^3
\label{M3}
\enq
Hence $\phi_r$ can be written as:
\beq
 \phi_r =  \frac{G M}{2 R_r^2} r \ \chi, \qquad \chi \equiv  \frac{1}{\Lambda} \int  \frac{R}{r} \tilde \rho (\tilde x') d^3 \tilde x'
\label{phir4}
\enq
$\chi$ is a dimensionless function that satisfies:
\beq
\lim_{r->\infty} \chi  = 1
\label{chi}
\enq
Similarly one can write the Newtonian potential as:
\beq
 \phi_N = - \frac{G M}{r} \ \psi, \qquad \psi \equiv  \frac{1}{\Lambda} \int  \frac{r}{R} \tilde \rho (\tilde x') d^3 \tilde x'
\label{phiN4}
\enq
$\psi$ is a dimensionless function that also satisfies:
\beq
\lim_{r->\infty} \psi  = 1
\label{psi}
\enq

\subsection{M33 Density Profile}

\begin{figure}
\includegraphics[width=0.9\columnwidth]{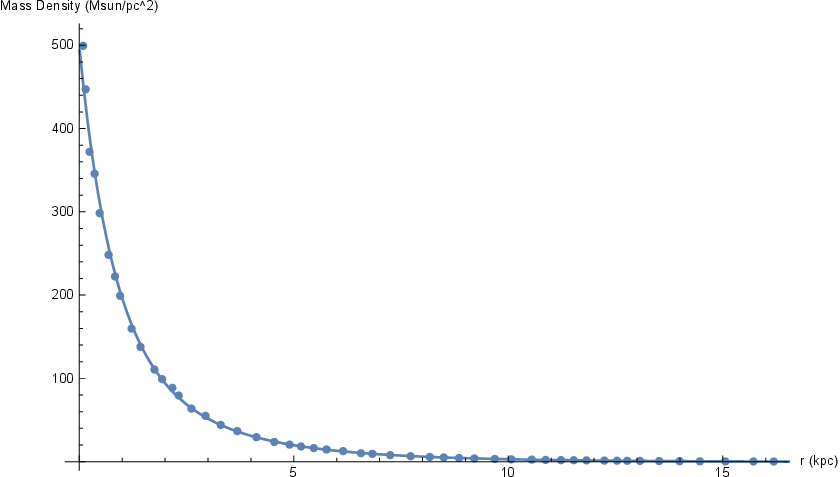}
 \caption{The M33 density radial distribution, the dots are data points \cite{Corbelli2} the solid line is a fit (\ern{kp}).}
 \label{radialdis}
\end{figure}

\begin{figure}
\includegraphics[width=0.9\columnwidth]{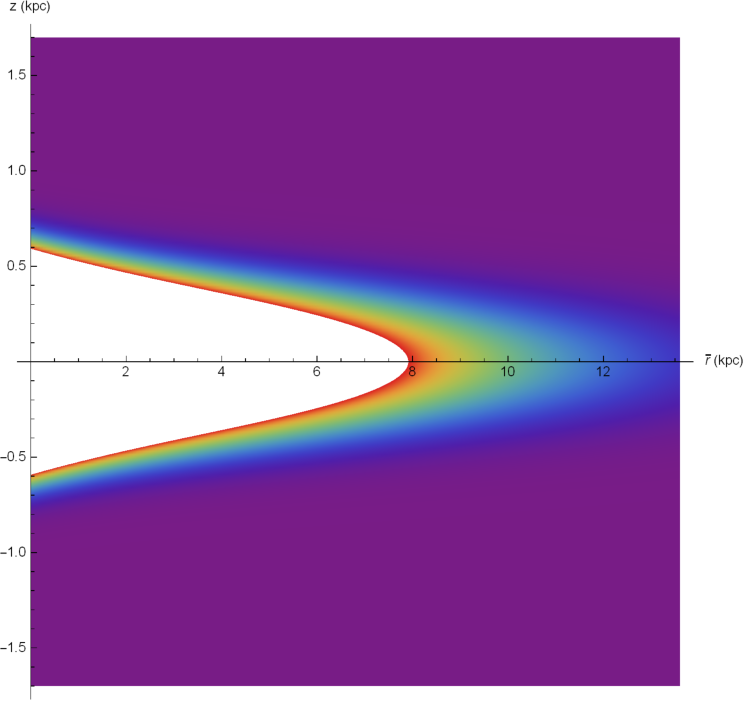}
 \caption{The M33 three dimensional mass density distribution model.}
 \label{3Ddis}
\end{figure}

In order to calculate the retardation and Newtonian gravitational potentials and hence the rotation curve through \ern{Eulersr2} we must know the density distribution in the galactic plane. This is usually done as follows; for the radial distribution we will use the luminosity distribution and assume a proportionality constant known as the mass to light ratio. Such radial density distribution is given in figure \ref{radialdis} which is based on the work of Corbelli \cite{Corbelli2} (see also Rega \& Vogel \cite{REV}) and its logarithm is fitted to a polynomial of order five as described in \ern{kp}:
\ber
\rho(\br) &\sim& e^{kp (\br)},
\nonumber \\
 kp (\br) &=& 6.21207 - 1.05618 \br + 0.137599 \br^2 - 0.0149017 \br^3
 \nonumber \\
 &+&  0.000865154 \br^4 - 0.0000209513 \br^5
\label{kp}
\enr
For the $z$ direction distribution orthogonal to the galactic plane we will assume a Gaussian profile with a typical width of $\sigma = 0.2$ kpc.
The mass distribution can be written as:
\beq
\rho(\br,z) = \rho_c  e^{kp (\br)} e^{-\frac{z^2}{\sigma^2}},
\label{dens}
\enq
in which we assume cylindrical symmetry. The three dimensional galactic mass distribution is depicted in figure \ref{3Ddis}.

\subsection{M33 Rotation Curve}

\begin{figure}
\includegraphics[width=0.9\columnwidth]{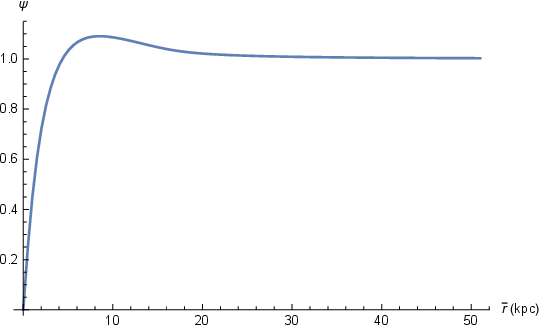}
 \caption{The $\psi$ function of M33, the function converges to $1$ for large distances as expected from \ern{psi}.}
 \label{psip}
\end{figure}

\begin{figure}
\includegraphics[width=0.9\columnwidth]{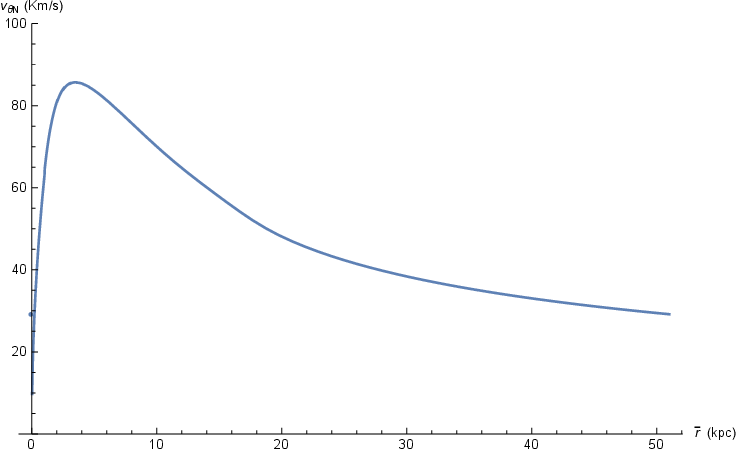}
 \caption{The Newtonian rotation curve of M33, the velocity decreases at large distances.}
 \label{vtN}
\end{figure}

\begin{figure}
\includegraphics[width=0.9\columnwidth]{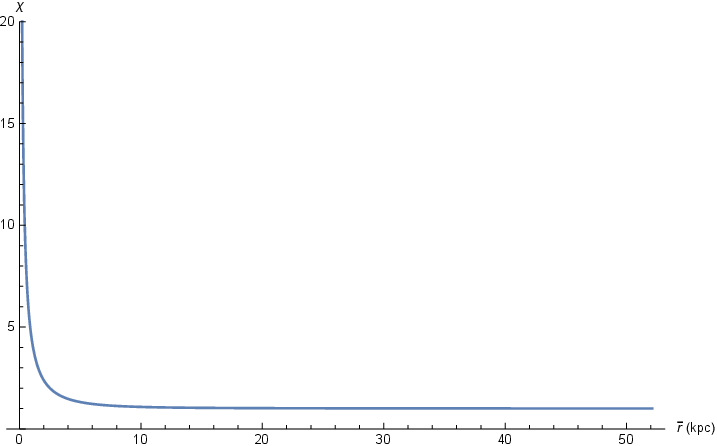}
 \caption{The $\chi$ function of M33, the function converges to $1$ for large distances as expected from \ern{chi}.}
 \label{chip}
\end{figure}

\begin{figure}
\includegraphics[width=0.9\columnwidth]{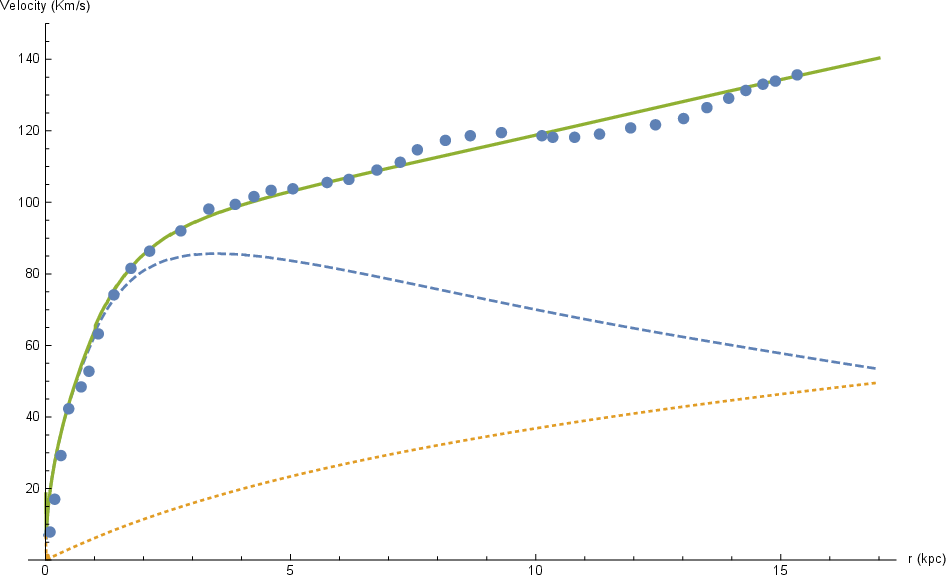}
 \caption{Rotation curve for M33. The observational points were supplied by Ms. Michal Wagman a PhD candidate in Ariel University under my supervision using \cite{Corbelli2}, the full line describe the complete rotation curve which is the sum of the dotted line describing the retardation contribution and the dashed line which is the Newtonian contribution.}
 \label{vcrhoc2}
\end{figure}

To obtain the rotation curve we need to evaluate the functions $\psi(\br)$ and $\chi(\br)$ which describe the non-trivial contribution of the retardation and Newtonian potentials respectively. We will assume for simplicity that the galaxy is cylindrically symmetric and that rotation curves are evaluated at the galactic plane. Thus according to \ern{phiN4} we need to evaluate numerically the integral:
\beq
\psi = \frac{1}{\Lambda} \int_{0}^{2 \pi} d \theta' \int_{0}^{\infty} d \br' \br'  \int_{-\infty}^{\infty} d z'
  \frac{\br}{R} \tilde \rho, \qquad  \tilde \rho = e^{kp (\br)} e^{-\frac{z^2}{\sigma^2}},
\label{psi2}
\enq
in which we use standard cylindrical coordinated $\br,\theta,z$. In the above:
\beq
\Lambda = \int_{0}^{2 \pi} d \theta' \int_{0}^{\infty} d \br' \br'  \int_{-\infty}^{\infty} d z'  \tilde \rho = 0.721007
\label{lambda3}
\enq
And:
\beq
 \frac{\br}{R} =\frac{1}{\sqrt{1 + \left(\frac{\br'}{\br}\right)^2  -   2 \left(\frac{\br'}{\br}\right) \cos(\theta') + \left(\frac{z'}{\br}\right)^2}}.
\label{Ror}
\enq
The result of the numerical evaluation of $\psi(\br)$ is given in figure \ref{psip}, we notice that due to the cylindrically symmetry it is enough to evaluate the function for the azimuthal angle $\theta=0$ as the result is similar for each azimuthal angle. As can be easily seen the $\psi$ function of M33 converges to $1$ for large distances as expected from \ern{psi}.
If we ignore the retardation potential contribution \ern{Eulersr3} then \ern{phiN4} will lead to the rotation curve depicted in figure \ref{vtN} which decreases for large distances. In the above we assumed a galactic (baryonic) mass of $10^{10}$ solar masses \cite{Corbelli2}.
As the true galactic rotation curve does not decrease we need to evaluate the retardation potential \ern{phir4} as well. For this
we evaluate the $\chi$ function as follows:
\beq
\chi \equiv  \frac{1}{\Lambda} \int_{0}^{2 \pi} d \theta' \int_{0}^{\infty} d \br' \br'  \int_{-\infty}^{\infty} d z'  \frac{R}{r} \tilde \rho
\label{chi2}
\enq
The result of the evaluation is given in figure \ref{chip} were it is obvious that the $\chi$ function approaches unity for large distances from the galactic center. The retardation potential and thus the retardation contribution to the velocity cannot be calculated without knowledge of the retardation distance $R_r$.
However, this can be obtained easily by fitting the observational Galactic rotation curve as demonstrated in figure \ref{vcrhoc2} which yields
a best fit for $R_r = 4.54$ kpc and retardation time of $t_r = 14,818.7$ years which are quite near to the rough estimates of section 5.

\section {Concluding remarks}

We show that "dark matter" and "MOND" effects are explained in the framework of
standard GR as effects due to retardation without assuming any exotic matter or modifications of the theory of gravity.
What will happen if the mass out side the galaxy is not depleted yet (or totally depleted)? In this case $\ddot{M} \simeq 0$
and retardation force should  vanish. This was indeed reported recently \citet{Dokkum} for the galaxy NGC1052-DF2.
Retardation effects in  electromagnetic theory were discussed in \citep{Tuval,YahalomT,Yahalom3}.
Finally we note that the same terms in the gravitation equation that are responsible for the gravitational radiation recently discovered is also responsible for the rotation curves of galaxies.
We regret that direct measurement of  the second temporal derivative of the galactic mass is not available. What is available is the remarkable fit between the retardation model and the galactic rotation curve as can be seen in figure 11, which constitute an indirect evidence for the galactic mass second derivative. The reader is reminded that competing theories like dark matter do not supply any observational evidence either. Despite the work of thousands of people and investment of Billions of Dollars there is still no evidence for dark matter. Occam's razor dictates that when two theories compete the one that makes less assumptions has the upper hand. In the case of retardation theory only baryonic matter is assumed and a large second temporal derivative of mass is related to density gradient caused by intergalactic gas depletion. There is no need to invent new kinds of matter that do not exist.
Nevertheless, this paper may serve as an inspiration for observing astronomers to better characterized the density gradients in galaxies and examine if the observations are indeed consistent with the theory.

The cuspy halo problem (also known as the core-cusp problem) refers to a discrepancy between the inferred dark matter density profiles of low-mass galaxies and the density profiles predicted by cosmological N-body simulations. Nearly all simulations form dark matter halos which have "cuspy" dark matter distributions, with density increasing steeply at small radii, while the rotation curves of most observed dwarf galaxies suggest that they have flat central dark matter density profiles ("cores"). This problem does not occur in the retardation model which denies the existence of dark matter. One cannot discuss flat or sharp profiles of dark matter if dark matter does not exist. The inherent problems with dark matters dynamics further strengthen the claim of this work that dark matter does not exist and the rotation curve characteristics attributed to dark matter should be attributed to retardation.

Finally the paper does not discuss dark matter in cosmological context this is left for future works.

\section*{Acknowledgement}

I would like to thank Ms. Michal Wagman a PhD candidate at Ariel University for supplying the data points of the rotation curve and the luminosity curve of the M33 galaxy.

\section*{Funding statement}

No funding received for this work.

\end {document}